\documentstyle[12pt]{article}
\input{psfig}

\begin{document}
\begin{flushright}
INR 0954/97 \\
September 1997 \\
hep-ph/9709304
\end{flushright}
\vspace*{3cm} 
\begin{center}
{\large \bf Jet Production in Regge-Limit of QCD \footnotemark \\}
\vspace*{1cm}
{\large Victor T. Kim${}^{\dagger}$ and 
Grigorii B. Pivovarov${}^{\ddagger}$ \\}
\vspace*{0.5cm}
{${}^\dagger$ : St.Petersburg Nuclear Physics Institute,
188350 Gatchina, Russia \\ 
${}^\ddagger$ : Institute for Nuclear
Research, 117312 Moscow, Russia} 

\vspace*{2cm}

{\large \bf Abstract}
\end{center}
Current issues on jet production in hadron 
collisions for Regge-limit of QCD are briefly discussed.

\footnotetext{presented by V.T.K. 
at the VIIth Blois Workshop
"Recent Advances in Hadron Physics", \\
\hspace*{0.55cm} Seoul, June 10-14, 1997, to appear in the Procceedings}

\newpage
\section{Introduction}

The energy reached in hadron
collisions and deep inelastic scattering provides
data \cite{CDF93,CDF96} on jet production whose understanding 
requires \cite{KP96d} an account of the resummation of the 
leading energy logarithms for QCD \cite{Kur76,Lip97}.
This motivates additionally efforts on
building an effective field theory
for Regge-limit of QCD. Among them are
reduction to two-dimensional theory \cite{Ver93}, 
analysis of symmetry properties of the effective theory \cite{Kir94,Lip95},
considerations in the large-$N_c$ limit \cite{Lip93},
with nonlocal Wilson-string operators \cite{Bal96}, 
with particular emphasis on the renormalization group \cite{KP97a}, and
with the renormalization group at high parton density 
\cite{Jal96}. These theories are presumably intended 
for any processes, exclusive or inclusive ones.
It turns out that one can build an economical effective
field theory for inclusive processes \cite{KP96c}.
In any case, for phenomenological implications one 
needs to involve somehow hadron structure (parton 
distrtibution functions
and/or wave functions). To this end
one needs to discuss a problem of factorization in hadron
collisions.

\section{Factorization Schemes for Regge-limit of QCD}

In perturbative QCD there are two important kinematical regimes:
hard scattering regime ($s \simeq -t = Q^2 \gg {\Lambda^2} $) and
Regge (semi-hard) regime  ($s \gg  -t = Q^2 \gg {\Lambda^2}$).
In hard scattering regime there are strong 
$k_{\perp}$-ordering: 
$k_{1,\perp} > k_{2,\perp} > ... >
k_{i,\perp} > k_{i+1,\perp} >...> k_{n,\perp}$
and locality in rapidity
$y_{1} \sim y_{2} \sim ... \sim
y_{i} \sim y_{i+1} \sim ... \sim y_{n}.$

\begin{figure}[htb] 
\vskip 7 cm
\includegraphics{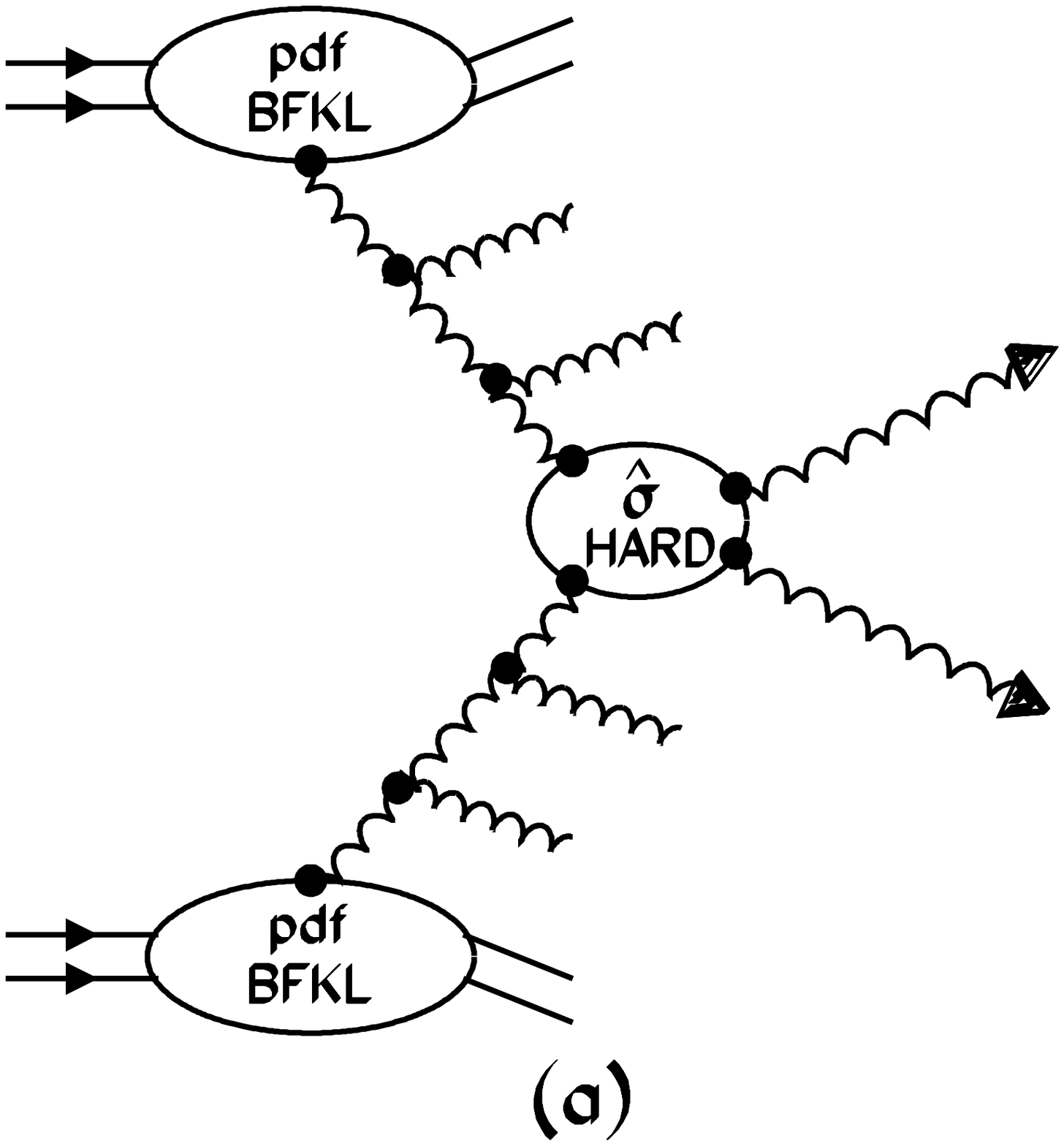} 
\vskip 0 cm
\includegraphics{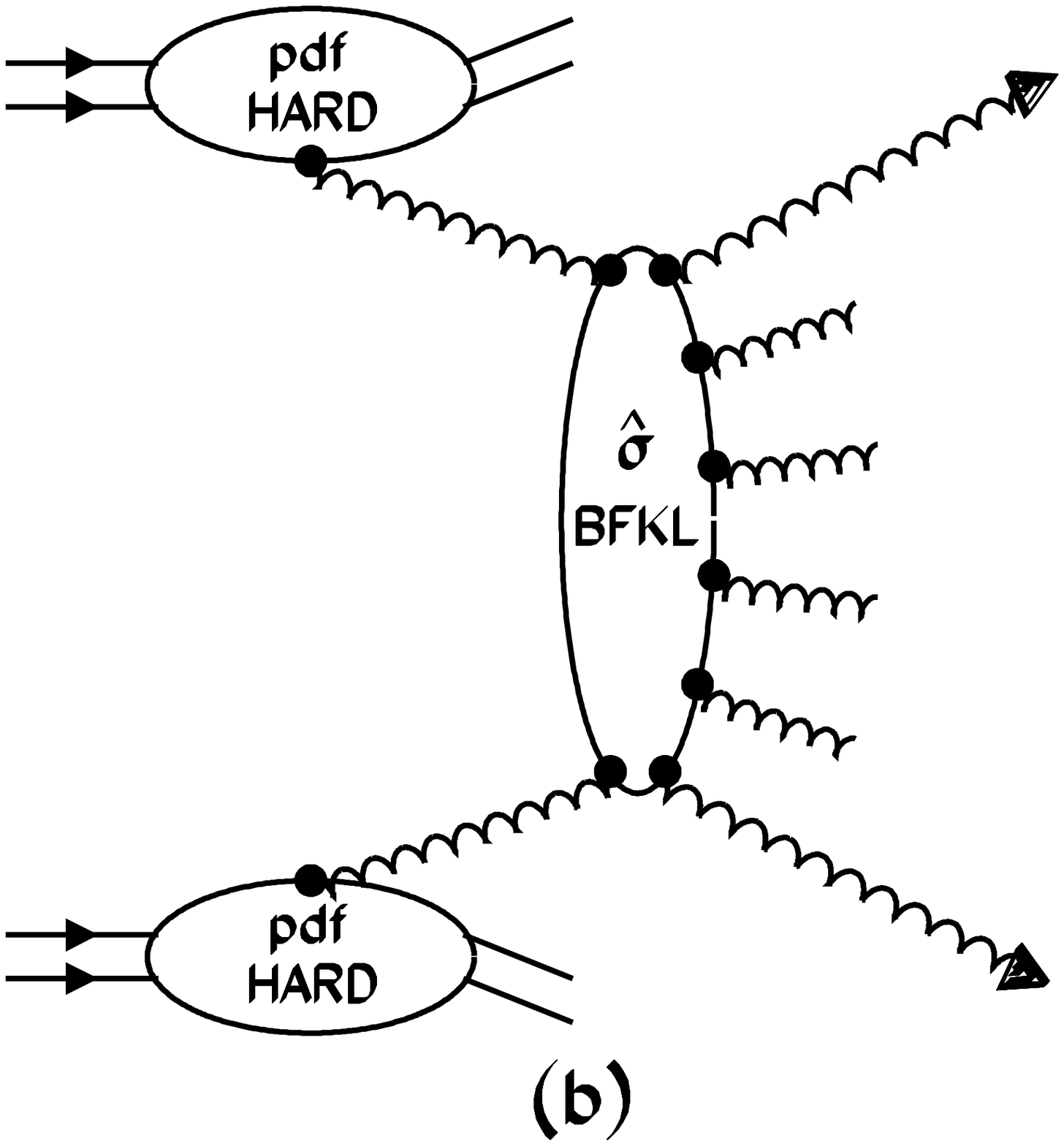} 
\vskip 7 cm
\includegraphics{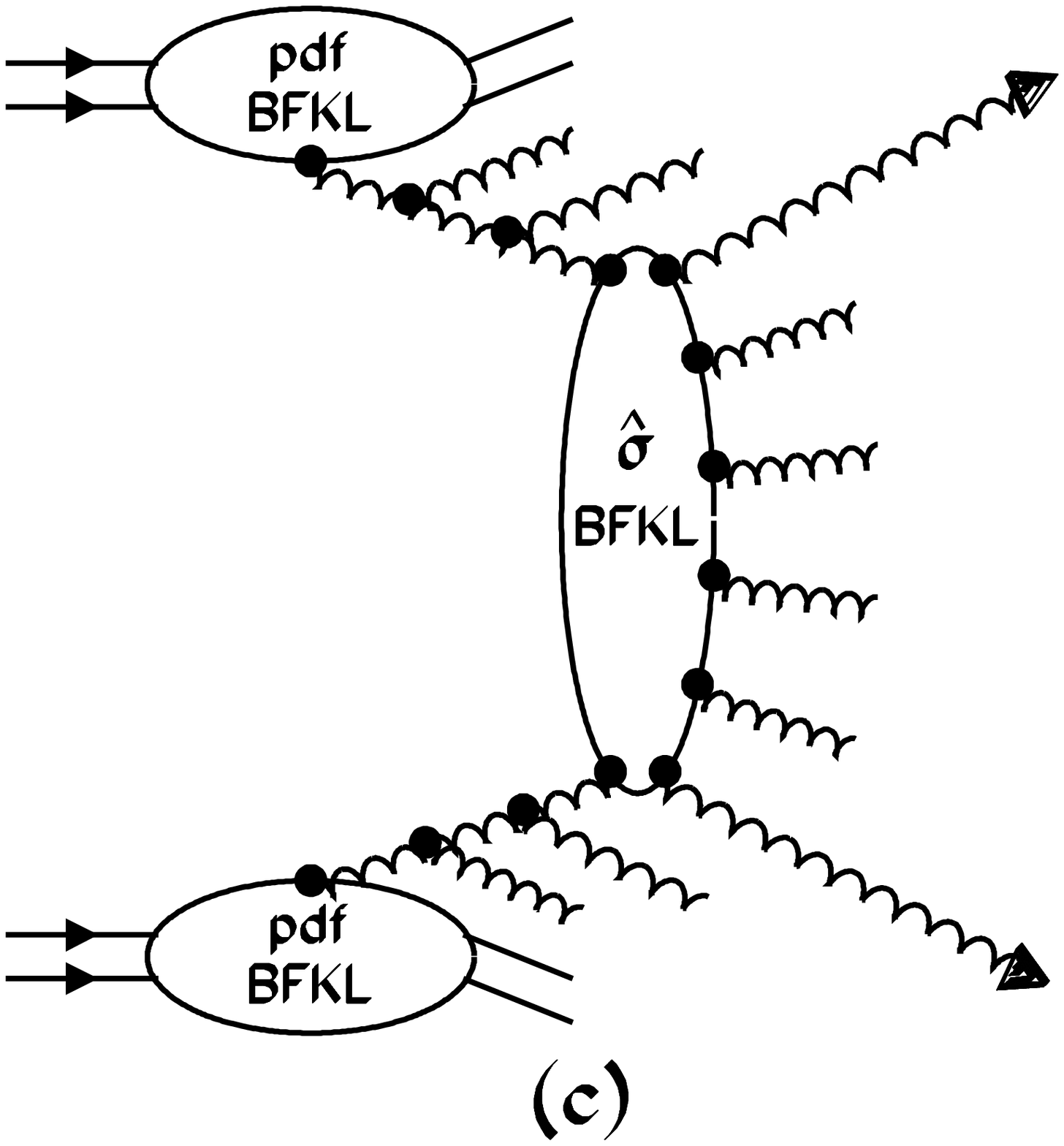}
\caption{(a) $k_{\perp}$-factorization; 
(b) Mueller-Navelet-factorization; (c) BFKL-factorization}
\end{figure}

In (multi-)Regge regime there is strong rapidity ordering:
$y_{1} > y_{2} > ... >
y_{i} > y_{i+1} >...> y_{n} $ and 
$k_{1,\perp} \sim k_{2,\perp} \sim ... \sim
k_{i,\perp} \sim k_{i+1,\perp} \sim ... \sim k_{n,\perp} $.
So, in semi-hard kinematics the BFKL subprocess is nonlocal
in rapidity space.

Unlike the hard scattering regime, where there is factorization 
theorem for inclusive processes (see \cite{Ste96} 
and references therein), in the Regge regime the factorization 
property is only a well-desired conjecture at the moment.

At present there are several factorization schemes 
for hadron collision in the Regge-limit of QCD.
One can divide it into three types which are transparent
in the case of dijet production in hadron collisions.
First one is $k_{\perp}$-factorization scheme \cite{Cat91} 
(Fig. 1a). $k_{\perp}$-factorization scheme allows to 
calculate processes in central region on rapidity, 
while second one, 
Mueller-Navelet(MN)-factorization scheme \cite{Mue87} 
(Fig. 1b), for processes
in most forward and backward regions which are separated
by large rapidity interval.
In other words, $k_{\perp}$-factorization scheme uses
BFKL-like parton distributions and hard subprocess (Fig. 1a),
and MN-factorization scheme uses BFKL-subprocess 
and hard parton distribution functions (Fig 1b). 
There are another scheme:
BFKL-factorization scheme \cite{KP96c,KP96b} (Fig. 1c)
which allows to calculate processes 
for any rapidity region. Within this scheme one can
calculate complete inclusive jet cross section, while
MN-factorization is dealing with selection of most 
forward/backward jets.

\section{Tests of BFKL Predictions in Hadron Collisions}

Azimuthal decorrelation in dijet production with most 
forward/backward jet selection calculated in \cite{Del94} 
is under study by D$\emptyset$ \cite{D096,D097} at the Fermilab Tevatron. 
Analysis with selection of most forward/backward 
jets seems to be not so promising since preliminary D$\emptyset$ 
data \cite{D097} and predictions of standard Monte Carlo 
(MC) generators, like Herwig and Pythia, 
as well as the new BFKL MC generators \cite{Sch97} 
are close to each other.

\begin{figure}[htb] 
\vskip 7 cm
\includegraphics{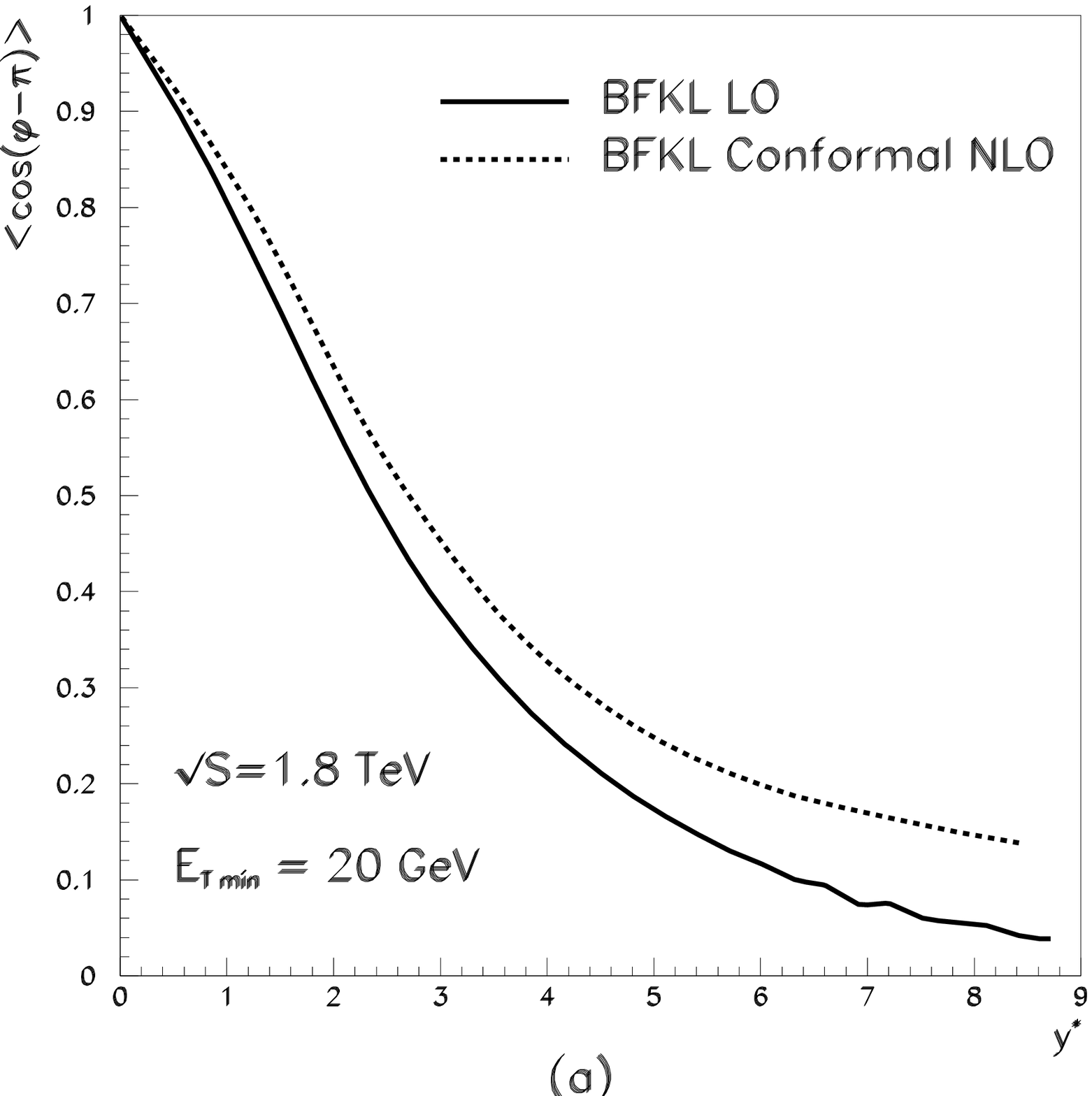}
\vskip 0 cm
\includegraphics{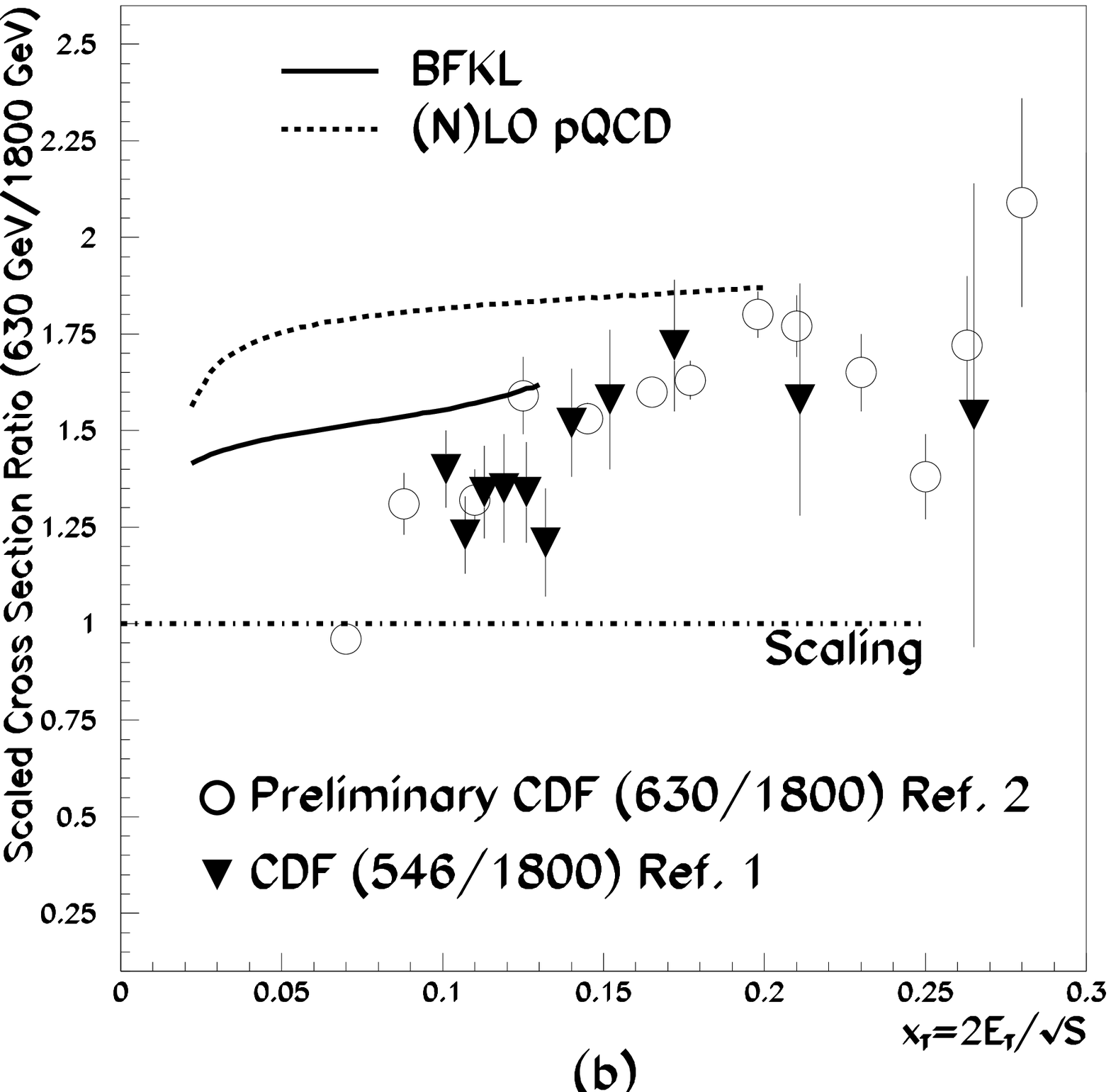}
\caption{(a) Inclusive dijet production: 
azimuthal decorrelation for BFKL LO $^{17,23}$ and 
for BFKL conformal NLO $^{23,24}$; 
(b) Inclusive single jet production: 
scaled cross section ratio $^{3}$}
\end{figure}

Futher progress for experimental tests of
BFKL predictions in hadron collisions should invoke
new observables sensitive to BFKL-effects \cite{Rys80}.
One of them can be azimuthal decorrelation of 
inclusive dijet production \cite{KP96b,KP96a} (Fig. 2a)
without most forward/backward jet selection.
Incorporation of complete next-to-leading order (NLO) \cite{Fad96} 
to BFKL predictions is desirable too.

\section{Discussion, Remarks and Conclusion}

Under these circumstances, it is crucial to have qualitative
predictions from the conventional QCD-improved parton model 
(without resummation of the energy logarithms) for 
the new kinematic domain. 
There is such prediction \cite{KPV97} which can be tested
at the Fermilab Tevatron, the CERN LHC \cite{Felix} and the VLHC 
\cite{VLHC} under discussion.
Namely, the QCD-improved parton model predicts that the ratio 
of scaled inclusive jet coss sections is not 
a monotonic function of its arguments, i.e. the inclusive 
jet production cross section, if measured in the natural units 
of the same cross section taken at another
(higher) energy of the collision, has extrema --- "dips" \cite{KPV97}.
If one does not observe the minima experimentally,
radical changes are motivated such as an alternative model 
of the elementary constituents inside the hadrons for semi-hard 
asymptotics. One example might be the color 
dipole model \cite{Mue94}.

To conclude, more deep understanding of factorization
property for Regge-limit of QCD is needed.
Experimental tests of BFKL-predictions and finding of 
good observables for BFKL-effects are very important for
studying of the new domain of QCD.

 We thank A.V. Efremov, S.Y. Jun, V.A. Kuzmin, 
 L.N. Lipatov, V.A. Matveev, H.S. Song, N. Varelas, J.P. Vary, 
 A.A. Vorobyov, H. Weerts and A.R. White for helpful discussions. 
 V.T.K. thanks the Blois Workshop Organizing Committee, 
 the Center Theoretical Physics of the Seoul 
 National University for warm hospitality and the 
 Korean Science and Engineering Foundation (KOSEF) for support.
 This work was partially supported by the Russian Foundation
 for Basic Research, grants No. 96-02-16717 and 96-02-18897.

\end{document}